# A novel reduced order model for vortex induced vibrations of long flexible cylinders

Giovanni Stabile[a,∗], Hermann G. Matthies[b], Claudio Borri[c]

[a]*SISSA Mathlab, International School for Advanced Studies, Via Bonomea 265, I-34136 Trieste, Italy*
[b]*Institute of Scientific Computing, Technische Universität Braunschweig, Mühlenpfordtstrasse 23 (8th floor), D-38106 Braunschweig, Germany*
[c]*Department of Civil and Environmental Engineering, University of Florence, Via di S. Marta 3, I-50139 Florence, Italy*

## Abstract

In this manuscript the development of a reduced order model for the analysis of long flexible cylinders in an offshore environment is proposed. In particular the focus is on the modelling of the vortex induced vibrations (VIV) and the aim is the development of a model capable of capturing both the in-line and cross-flow oscillations. The reduced order model is identified starting from the results of a high fidelity solver developed coupling together a Finite Element Solver (FEM) with a Computational Fluid Dynamics (CFD) solver. The high fidelity analyses are conducted on a reduced domain size representing a small section of the long cylinder, which is nevertheless, already flexible. The section is forced using a motion which matches the expected motion in full scale, and the results are used for the system-parameter identification of the reduced order model. The reduced order model is identified by using a system and parameter identification approach. The final proposed model consists in the combination of a forced van der Pol oscillator, to model the cross-flow forces, and a linear state-space model, to model the in-line forces. The model is applied to study a full scale flexible model and the results are validated by using experiments conducted on a flexible riser inside a towing tank.

## 1. Introduction

Due to the gradual depletion of oil and gas resources onshore and in shallow waters, recent years have seen an increasing interest in deeper waters, where a large proportion of the remaining oil and gas is located. The recent interest in deeper waters comes not only from the petroleum industry but also from the renewable energy sector. The sea, especially in deep waters, has a huge energy potential, which could be exploited using wave energy converters, solar power plants and wind turbines located on offshore platforms. Long slender cylinders are found in many offshore applications and are the representative system for mooring lines, risers, umbilicals and free spanning pipelines in deep water. The response of these kinds of structures to wave, current and tide loads may be complex, and phenomena such as vortex induced vibrations, unsteady lock-in, dual resonance, and travelling waves response may occur [62]. Much progress has been made to understand the hydrodynamic forces that have to be used for these structures but an efficient and reliable model, especially dealing with vortex induced vibrations [44], is still missing in literature. Computational Fluid Dynamics (CFD) methods have been demonstrated to be a possible way of getting the response of flexible structures in an offshore environment, especially considering the increasing of the available computational power, but they are still not applicable to long-term simulations and to values of the Reynolds number interesting for practical applications.

Although in the future CFD methods will probably be the first choice for design purposes, at the moment we still have to rely on simplified and approximated methods mostly based on experimental investigations conducted on rigid cylinders undergoing forced or free vibrations [44, 61, 60]. This contribution aims to develop a new time domain simplified method identified through high-fidelity FSI simulations conducted on flexible cylinders undergoing forced oscillations.

The paper is organised as follows: in § 2 a brief overview regarding VIV phenomena and existing methods for their simulations is presented, in § 3 the methods used for the evaluation of the reduced order method are introduced. § 4 describes the fluid-structure interaction solver developed and used for the high fidelity analyses. In § 5 the ROM proposed in this manuscript is described and in § 6 numerical tests are presented and discussed. Finally, in § 7, some conclusions are drawn and perspectives for future works are provided.

---

∗Principal corresponding author
  *Email addresses:* `gstabile@sissa.it` (Giovanni Stabile), `wire@tu-bs.de` (Hermann G. Matthies), `cborri@dicea.unifi.it` (Claudio Borri)

## 2. Brief overview on the VIV phenomenon

Long flexible cylinders in an offshore environment are often exposed to the phenomenon of vortex-induced vibrations (VIV). VIV are an important phenomenon in many different engineering fields and have been studied in either air or water for many years. The content of this section is summarized from the numerous reviews on the topic available in literature [60, 61, 44, 62, 16]. This phenomenon is caused by the oscillating flow arising from the alternate vortex shedding. Among all possible existing phenomena, that may happen on flexible cylindrical structures in an offshore environment, VIV are one of the most dangerous and hard to predict. If a rigid and fixed cylinder is considered, the frequency of the vortex shedding phenomenon follows the Strouhal law [50]:

$$St = \frac{f_v D}{U} \qquad (1)$$

in which $St$ is the Strouhal number, $f_v$ is the frequency of the vortex shedding, and $U$ is the free stream velocity.
For a rigid and fixed circular cylinder, the Strouhal number, over a certain range of the Reynolds numbers, assumes a constant value approximately equal to 0.2. In this range of values, which are particularly interesting for practical applications, the relationship between the free stream velocity and frequency of oscillation is linear. Conversely, if it is considered a cylinder which is free to vibrate or forced to move the phenomenon does no longer obey the Strohual law [50]. When the vortex shedding frequency approaches the natural frequency the so called lock-in phenomenon is observed. The synchronisation of the vortex shedding frequency to the frequency of oscillation occurs over a certain range of the flow velocity. In this range the vortex shedding phenomenon is driven by the frequency of oscillation of the cylinder.

### 2.1. Literature survey on VIV modelling

Although a lot of research has been performed in this field, three basic different methods to predict the behaviour of a slender cylinder subjected to VIV can be found:

- Semi-empirical models
- Navier-Stokes models
- Simplified wake models

Here the methods are only briefly introduced and discussed, a comprehensive review regarding these different approaches and a comparison of their results can be found in [10].

#### 2.1.1. Semi-empirical models
These methods are widely used by numerous authors and are nowadays the standard for many commercial codes used in offshore engineering such as VIVA [52, 63], VIVANA [33, 34], SHEAR7 [54, 55]. The instantaneous amplitude of oscillation is evaluated using appropriate force coefficients. These are evaluated on experimental tests on rigid cylinders undergoing free or forced vibrations. In these approaches it is normally assumed that the cylinder is oscillating only in cross-flow direction and when also the in-line oscillation is considered it is assumed to be decoupled respect to the cross-flow motion. Within this method, natural frequencies of the structure are evaluated and than the modes which are most likely to be excited by the vortex shedding are identified. Normally only the steady in-line response of the structure is identified.

#### 2.1.2. CFD models
In CFD methods the flow field around the cylinder is computed by solving numerically the unsteady Navier-Stokes equations. [5, 4] proposed a full three dimensional analysis of a long flexible cylinder. In these works the authors coupled a structural code with a CFD solver with direct numerical simulation of three dimensional unsteady Navier-Stokes equations, and the structural response is investigated. [7] used a geometrically-exact beam model for the coupling with a CFD code which solves the unsteady incompressible Reynolds-averaged Navier-Stokes equations enabling vortex induced vibration configurations to be handled. In his PhD thesis, [22] investigated the vortex shedding on a riser. He used a direct FEM integration solver for the structural dynamic and an unsteady incompressible Navier-Stokes solver with a LES turbulence model for the fluid dynamic. A complete review on numerical methods for VIV is also presented in his thesis. [1] performed several high-fidelity CFD analyses with Navier-Stokes DNS methods in order to represent the flow field with few dominant modes using a POD. The emphasis is here given to the control of VIV using fluid actuators. [17] applied POD, after several high-fidelity CFD analyses, in order to condense the information contained in the flow field into a limited set of modes. In [48, 49] some of the authors of the present paper applied POD for parametrized model reduction of vortex shedding with a particular focus on pressure stabilisation at reduced order level. When summarising the whole of the cited works, CFD methods can be classified into two different classes:

- **Full 3D simulations** where the flow is discretized using 3D elements [4, 6, 5, 20]. This approach has the advantage of completely capturing the three-dimensionality of the flow, but even with todays computational resources this is too demanding for the case of long cylinders and for flows with realistic values of the Reynolds number.

- **Strip theories** where the flow is modelled using bidimensional analysis at several different positions along the cable length. The flow at different planes is completely independent [32, 19, 45, 58, 59] and the three-dimensionality is only due to the structural model.

This approach is computationally less demanding, and flow with realistic values of Reynolds number can be simulated. The disadvantage is that the three-dimensionality of the flow cannot be captured and forces at different planes have to be interpolated, and it is not easy to find a general way to perform such an interpolation.

DNS CFD methods, even if have been used, are not suitable for riser VIV simulation. Contributions where DNS methods are used always consider a small Reynolds number which is not compatible with real conditions. RANS and LES CFD methods demonstrated to be suitable to study VIV effects on short rigid cylinders, even for higher Reynolds numbers, but in case of long flexible cylinders they are still computationally too demanding and for this reason they cannot be used for design purposes where a large number of simulations are needed. The analyses are performed for a relatively small value of the Reynolds number or for small value of the length ratio. Moreover, these approaches require high computational resources and long computational times which make theme unsuitable for design purposes. To give an idea of the computational cost of a full 3D-dimensional simulation, we refer here to [5]. They conducted a DNS CFD analysis on a long flexible cylinder ($L/D = 200$) at $Re = 1100$ on a 512 cores computer. The simulation time for each time step resulted to be approximately equal to 3s and considering that a cross-flow vibration cycle counted 13000 time steps this would led to a simulation time approximately equal to 11h for each vibration cycle.

2.1.3. Simplified wake models

These models, instead of completely solving the flow field, model the flow forces using simplified models. The 3-dimensionality of the problem is reconstructed using a strip theory. Most of the simplified model present in literature are able to predict only the hydrodynamic forces for the cross-flow direction [36]. The fluctuating value of the drag is normally not considered and when taken into account is supposed to be independent from the lift coefficient [15]. Large parts of the proposed models are based on the idea of the wake oscillator [43]. In this approach, the dynamics of the wake behind the cylinder is modelled by one single variable which is supposed to satisfy a non-linear differential equation which is self-excited and self-limited. For this purpose, systems normally used to model vortex shedding on static cylinders, such as the Rayleigh (Eq. 2) [18] or the van der Pol (Eq. 3) equations, are extended [15, 47] with a forcing term depending on the motion of the cylinders:

$$\ddot{L}_c + \omega_s^2 l - \mu_r \dot{L}_c + \alpha_r \dot{L}_c^3 = F(d_s, \dot{d}_s, \ddot{d}_s) \quad (2)$$
$$\ddot{L}_c + \omega_s^2 l - \mu_v \dot{L}_c + \alpha_v L_c^2 \dot{L}_c = F(d_s, \dot{d}_s, \ddot{d}_s) \quad (3)$$

in which $L_c$ is the lift coefficient, $\omega_s$ is the vortex shedding frequency for a stationary cylinder, $\mu_r$, $\mu_v$ and $\alpha_r$, $\alpha_v$ are the linear and non-linear damping terms identified respectively for the Rayleigh and the van der Pol equation, and $F$ denotes a forcing term which couples the structural displacement $d_s$ or eventually its derivatives $\dot{d}_s$, and the wake oscillator equation. This forcing term differs among different authors but it usually depends linearly on acceleration, velocity or displacement along the considered direction. [18] proposed a Rayleigh wake oscillator. The structure is represented by a 1-DOF elastically mounted rigid cylinder, modelled as a linear damped second order dynamical system. The coupling term is in this case proportional to the transverse cylinder velocity:

$$\ddot{L}_c + \omega_0^2 L_c - \mu\omega_0 \dot{L}_c + \frac{\alpha}{\omega_0}\dot{L}_c^3 = B\dot{d}_{CF} \quad (4)$$

[15] coupled a 1-DOF structural oscillator, free to vibrate only along the cross flow direction, with a van der Pol wake oscillator. In his work different coupling terms are analysed and it is assessed that the configuration which gives the best results is the acceleration coupling model. Less research has been performed concerning the in-line displacements and the hydrodynamic forces for the in-line direction. [38] modelled the lift coefficient using a standard van der Pol oscillator and for the drag, considering that the fluctuating frequency of hydrodynamic forces along the in-line direction is twice the cross-flow frequency, used a quadratic coupling between lift and drag in a Reynolds decomposition setting:

$$D_c = D_{c,m} - KL_c\dot{L}_c \quad (5)$$

in which $K$ is a coefficient that must be determined experimentally and $D_c$ and $D_{c,m}$ are respectevely the total and fluctuating drag coefficients. [47] proposed to model both the cross-flow and in-line fluctuating hydrodynamic forces using a van der Pol wake oscillator. The structure is modelled with a 2-DOF double Duffing oscillator. For the forcing of the van der Pol oscillator they used the structural transversal acceleration for the cross-flow direction and the acceleration along the direction of the flow for the in-line direction. [36] proposed a model allowed to capture the chaotic response of marine structure under VIV. For each cross-section recent velocity history is compressed using Laguerre polynomials. The instantaneous velocity is used inside an interpolation function carried out using experimental investigations. The recent trajectory is described so that it is more precise for the immediate past and becomes less precise further back in time.

3. Introduction to the present method

The method proposed in this paper aims to exploit both the advantages of CFD methods and of simplified wake models. The response of a piece of flexible cylinder is studied only locally, on a reduced domain size. The limited size of the domain permits to use a complete three-dimensional analysis with values of the Reynolds number suitable for practical applications. The logical process on which the present method is based is explained in Figure 1. The local model of the cable, henceforth called the fine-scale model or the high-fidelity model, is subjected to an imposed motion which matches the statistical properties of the expected motion in full scale. The hydrodynamic forces along the cylinder of this model are measured. The results of these high-fidelity simulations together with the correspondent input motion can be used to feed a system-parameter identification technique which permits the definition of a reduced order model. In a such a manner the computational effort can be divided into two different phases an on-line and off-line one. During the off-line phase the high fidelity simulation are performed and the parameters of the reduced order model are evaluated. These can be directly used during the on-line phase to simulate the overall behaviour of the structure. The resulting developed model can be added to any computational structural dynamic solver. More details about the development of the ROM are given in § 5.

## 4. The High Fidelity Model

Since in this manuscript the identification of the reduced order model is based on the results of numerical simulations, the method presented in § 3 is strongly based on the accuracy of the used fluid-structure interaction solver. The interaction between the structure and the fluid is a coupled problem and in the next sections it will be explained how it is solved in the context of this work. The solid problem is solved using a finite element approach [64, 24] while the fluid problem is solved using a finite volume approach [57, 3]. The coupled problem is solved here using a partitioned strategy with a so called direct force-motion transfer (DFMT) method [42]. In general, a fluid-structure interaction problem is formed by three different sub-problems [9, 8]: the fluid problem, the solid problem, and the mesh motion problem. The structural problem will be henceforth indicated with $s$ while the fluid problem will be indicated with $f$.

### The mathematical model of the solid problem

The solid problem is governed, in an Eulerian frame, by the momentum balance equation in terms of Cauchy stresses:

$$\boldsymbol{\nabla} \cdot \boldsymbol{\sigma} + \rho_s(\boldsymbol{b} - \ddot{\boldsymbol{u}}) = \boldsymbol{0} \text{ in } \Omega_s \times [0,T] \quad (6)$$

in which $\boldsymbol{\sigma}$ is the Cauchy stress tensor, $\rho_s$ is the solid density, $\boldsymbol{b}$ is the body force vector, $\ddot{\boldsymbol{u}}$ is the acceleration vector. $\Omega_s$ is the structural domain, and $T$ is the length of the considered time window.

### The mathematical model of the fluid problem

The water is modelled as a Newtonian, incompressible, viscous, isothermal and isotropic flow. Since the fluid domain is not static but changing in time due to the deformation of the solid body one needs to reformulate the Navier-Stokes equation for incompressible and viscous flows considering the motion of the FSI interface. This is done using an arbitrary-Lagrangian-Eulerian (ALE) [13, 14] framework referenced to a frame moving with a velocity $\boldsymbol{v_m}$. One needs to replace the convective term inside the velocity $\boldsymbol{v}$ with the convective velocity $\boldsymbol{v_c} = \boldsymbol{v} - \boldsymbol{v_m}$, where $\boldsymbol{v_m}$ is the velocity of the moving part of the domain The momentum balance equation and the continuity equation, for the fluid domain, can be written in an Eulerian frame as:

$$\frac{\partial \boldsymbol{v}}{\partial t} + (\boldsymbol{v_c} \cdot \boldsymbol{\nabla})\boldsymbol{v} - \nu \boldsymbol{\nabla}^2 \boldsymbol{v} = -\frac{1}{\rho_f} \boldsymbol{\nabla} p \quad \text{in } \Omega_f \times [0,T]$$
$$\boldsymbol{\nabla} \cdot \boldsymbol{v} = 0 \quad \text{in } \Omega_f \times [0,T]$$

in which $\boldsymbol{v}$ is the flow velocity vector, $t$ is the time, $\nu$ is the flow viscosity, and $p$ is the pressure. $\Omega_f$ is the fluid domain, and $T$ is the length of the considered time window. In the present work the solid and the fluid problem are solved using different space discretization techniques. The solid problem is solved using finite elements [64, 24], while the fluid problem is solved using finite volumes [57, 3].

### The coupling strategy and the mesh motion problem

The mesh motion problem is solved imposing the matching of velocities and stresses at the FSI interface:

$$\boldsymbol{v_s} = \boldsymbol{v_f} \text{ on } \partial\Omega_{FSI} \times [0,T] \quad (8)$$
$$\boldsymbol{n} \cdot \boldsymbol{\sigma} = -p\boldsymbol{n} + 2\nu(\boldsymbol{n} \cdot \boldsymbol{\nabla}^S)\boldsymbol{v_f} \text{ on } \partial\Omega_{FSI} \times [0,T]$$

where $\boldsymbol{n}$ is the vector normal to the FSI-interface and, $\boldsymbol{\nabla}^S$ is the symmetric part of the gradient operator. The matching of variables at the interface is enforced with an implicit scheme which conserves the energy at the interface. For each time step of the coupled simulation $t_{c,i}$ an iterative cycle on the velocity residual is performed until the achievement of a desired tolerance:

$$\boldsymbol{res} = \boldsymbol{v_{s,i}} - \boldsymbol{v_{f,i}} \leq TOL \quad (9)$$

The iterative scheme is realized using a block Gauss-Seidel procedure reported in Algorithm 1 [37]. In the algorithm $T_0$ is the initial time, $T$ is the simulation time, $\Delta t_f$, $\Delta t_s$ and $\Delta t_c$ are the time step of the fluid, solid and coupled simulation respectively, $TOL$ is the tolerance that defines the maximum allowed value of the residuals $\boldsymbol{res}_{N+1}^{(k)}$. In the algorithm the subscript defines the $N^{th}$ time step while the superscript defines the $k^{th}$ sub-iteration. The starting point of the algorithm is the prediction of the fluid velocity $\boldsymbol{V}_{f,N+1}^{(0)}$ at the FSI interface based on the previous converged values of the velocity at the interface $\boldsymbol{V}_N^{(k_{max})}, \boldsymbol{V}_{N-1}^{(k_{max})}$, etc. For values of the subiteration index $k > 0$ the Aitken's relaxation factor $\omega$ is evaluated using the expression reported in the algorithm. The fluid problem is then solved and the fluid forces $\boldsymbol{F}_{f,N+1}^{(k)}$ acting on the FSI interface are determined. These are applied on the structure and the solid problem is solved in order to determine the velocity of the solid points at the interface $V_{S,N+1}^{(k)}$. The residual $res_{N+1}^{(k)}$ is evaluated as the difference between the solid and fluid velocities at the interface. Finally the predicted value of the fluid velocity at the interface is updated using the new determined solid velocity at the interface and the evaluated Aitken's relaxation factor. The time step size of the fluid sub-problem may differ the time step size of the solid sub-problem. The only requirement is that each sub-problem has to be performed the sufficient number of times necessary to reach the time step size of the coupled simulation. Inside the algorithm the term $\omega$ is a scalar value obtained by Aitken's relaxation [31]. The Aitken's relaxation factor is used to speed up the computation but the coupling algorithm also works with a constant relaxation value. This procedure works with any CFD solver and any CSD solver which can be coupled properly; here the solid sub-problem is solved using the FEM solver FEAP [51], while the fluid sub-problem is solved using the FVM solver OpenFOAM [25]. The coupling is realized using the approach of software components and the component template library (CTL) [37] is used as common middleware. Here the features of the FSI solver are only briefly reported: for additional details please refer to [28, 29]. As already mentioned in this work, the solid and the fluid problems are solved independently and then coupled together using the explained coupling algorithm.

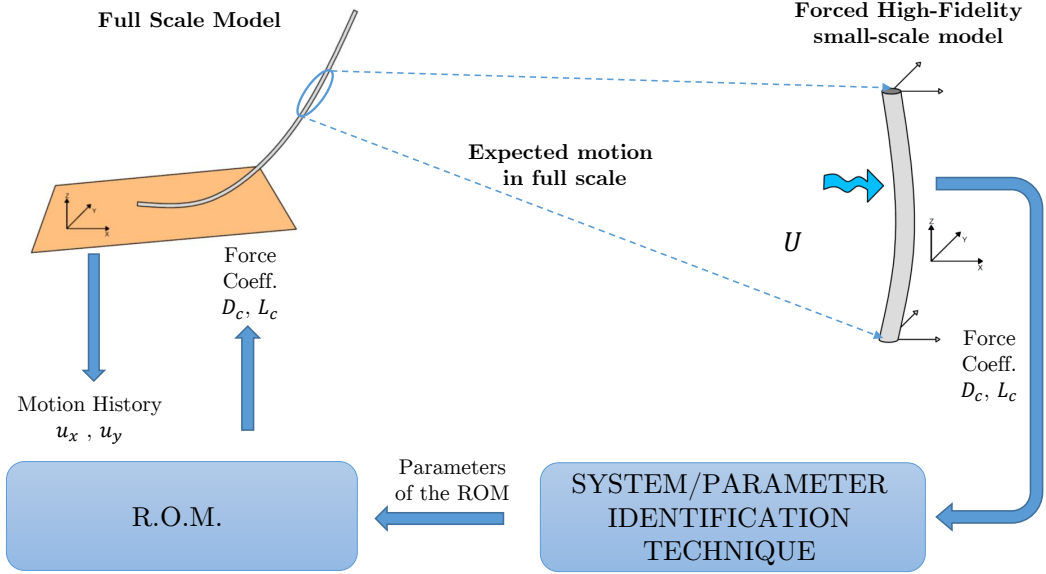

Figure 1: Flowchart of the proposed method

### 4.1. Setting of the fluid problem and its validation

The computational domain is represented in Figure 2 and is structured using a polar distributed grid in the proximity of the cylinder and a Cartesian distributed grid in the other regions. Only hexahedral elements are used. The region near to the wall is refined in order to obtain a $y^+$ number close to the unity. An equally spaced mesh is used along the vertical direction ($\Delta z \approx D/2$). The distance of the cylinder from the inlet is equal to $8D$, the distance from the outlet is equal to $15D$, and the domain width is equal to $20D$. The domain height is equal to $40D$. The mesh counts a total number of 360960 cells. The cylinder is divided into 80 equispaced elements along its circumference. The central part, which is discretized using a polar distribution, counts 45 cells along the radial direction with a cell expansion ratio equal to 3. The base mesh, which is external to the refined zone, is discretised with 32 cells along the vertical direction ($y$) and with 41 cells along the horizontal direction ($x$). For what concerns the numerical schemes a backward Euler scheme is used for the time integration, a second order linear upwind scheme is used for the discretization of the convective term, a central differencing scheme is used for the discretizazion of the diffusive term and for the discretization of the gradient of pressure. The length of the small-scale is chosen in relation with the correlation length of the fluid forces along the cylinder length. In fact to capture completely the three-dimensional effect the computational domain should be at least longer than the maximum correlation lenght. In lock-in conditions the correlation length is increased but normally it is always lower than 40 diameters as discussed in [21]. The flow has has a constant uniform velocity at the inlet and constant zero pressure value at the outlet. Sides are modelled with slip conditions. The PIMPLE algorithm [25] and an Euler implicit time integration scheme are used. A constant time step $\Delta t_f$. The time step is chosen a priori, before the beginning of each simulation, making a check on the maximum Courant number running a simulation on a static case. The check on the time step is performed in order

**Algorithm 1** Coupling algorithm
---
Given: initial time $T_0$, length of the simulation $T$, time step size of the fluid simulation $\Delta t_f$, time step size of the solid simulation $\Delta t_s$, time step size of the coupled simulation $\Delta t_c$, the tolerance $TOL$

  **while** $t < T$ **do**
    $k = 0, \omega^{(0)} = \omega_0$
    **while** $res_{N+1}^{(k)} < TOL$ **do**
      **if** $k = 0$ **then**
        Predict fluid v. at the inter.: $\boldsymbol{V}_{f,N+1}^{(0)} = P(\boldsymbol{V}_N^{(k_{max})}, \boldsymbol{V}_{N-1}^{(k_{max})}, \ldots)$
      **else**
        $\omega^{(k)} = -\omega^{(k-1)} \frac{res_{N+1}^{(k-1)} \cdot (res_{N+1}^{(k)} - res_{N+1}^{(k-1)})}{\|res_{N+1}^{(k)} - res_{N+1}^{(k-1)}\|^2}$
      **end if**
      Given $\boldsymbol{V}_{f,N+1}^{(k)}$ solve the fluid problem in ALE formulation $\to \boldsymbol{F}_{f,N+1}^{(k)}$
      Given $\boldsymbol{F}_{f,N+1}^{(k)}$ solve the solid problem $\to \boldsymbol{V}_{S,N+1}^{(k)}$
      Evaluate residual $\boldsymbol{res}_{N+1}^{(k)} = \boldsymbol{V}_{S,N+1}^{(k)} - \boldsymbol{V}_{f,N+1}^{(k)}$
      Update fluid velocity at the interface $\boldsymbol{V}_{f,N+1}^{(k+1)} = \boldsymbol{V}_{S,N+1}^{(k)} + \omega^{(k)} \boldsymbol{res}_{N+1}^{(k)}$
      $k = k + 1$;
    **end while**
    $N = N + 1, t = t + \Delta t_c$
  **end while**

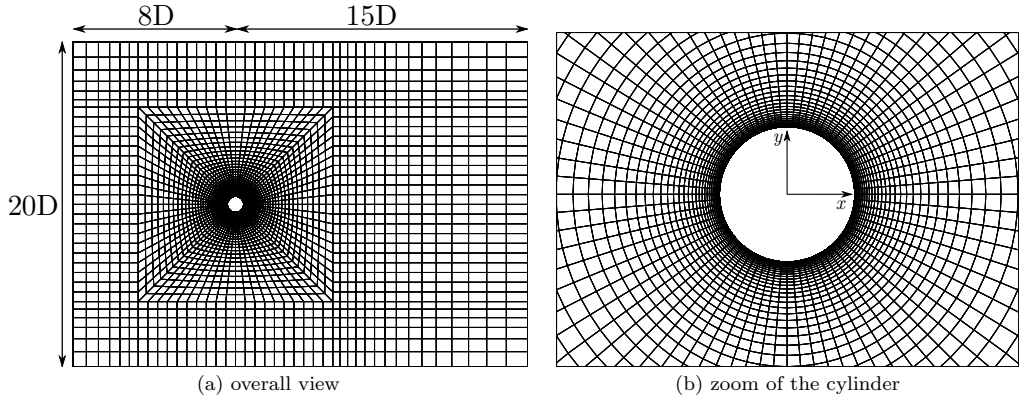

Figure 2: The left picture (a) shows a sketch of the structured computational grid used for the high fidelity simulations together with the main dimensions of the computational domain $\Omega_f$, as a function of the cylinder diameter $D = 0.102$m. The right picture (b) reports a zoom of the mesh in proximity of the cylinder.

|   | RMS LC | MEAN DC | MEAN LC | St |
|---|---|---|---|---|
| Exp. Value | 0.24 | 1.16 | 0.01 | 0.19 |
| Present | 0.33 | 1.25 | 0.0036 | 0.20 |
| A | 0.53 | 0.70 | 0.000 | 0.28 |
| B | 0.47 | 0.87 | 0.000 | 0.25 |
| C | 0.83 | 1.05 | -0.002 | 0.25 |
| D | 0.60 | 1.37 | 0.038 | 0.18 |
| E | 0.18 | 1.10 | -0.030 | 0.28 |
| F | 0.51 | 1.49 | 0.000 | 0.20 |
| G | 0.86 | 1.14 | 0.000 | 0.24 |
| H1 | 1.08 | 1.28 | 0.130 | 0.29 |
| H2 | 0.10 | 0.54 | 0.010 | 0.35 |
| H3 | 0.58 | 1.38 | -0.040 | 0.22 |
| H4 | 0.99 | 1.70 | -0.010 | 0.21 |

Table 1: Comparison with reference values coming from numerical and experimental results. The reference values are taken from [41]. The comparison is reported for $Re = 6.31 \times 10^4$. In the table are reported the values of the Root Mean Square of the Lift Coefficient (RMS LC), the mean values of the Drag and Lift Coefficients (MEAN DC, MEAN LC) and the Strouhal number (St).

to obtain a maximum Courant number which is always smaller than 1. In order to validate the whole model, the first step is to validate the small scale model and the CFD solver. A CFD simulation on a static domain is performed and the results are compared to numerical and experimental data in terms of lift, drag, and Strouhal number. Lift and drag forces are described in terms of the drag coefficient and lift coefficients:

$$D_c = \frac{F_D}{0.5\rho U^2 D} \;,\; L_c = \frac{F_L}{0.5\rho U^2 D} \;, \qquad (10)$$

where $F_D$ and $F_L$ are the drag and lift force per unit of length respectively and $U$ is the free stream velocity of the flow. In Figure 3 the time evolution of the lift and drag coefficient is reported. All the results are carried out using a Reynolds number $Re = 6.31 \times 10^4$ which matches the Reynolds value of the benchmark from [41]. The results are compared to the same benchmark case in terms of root mean square (RMS) of the lift coefficient, mean drag coefficient, and Strouhal number. As one can see from Table 1, results match well with experimental and previous numerical results.

*4.2. Setting of the structural problem*

The structure is discretised with 40 equally spaced geometrically-exact beam elements with the same spacing used for the fluid sub-problem. The same time step size of the fluid sub-problem is used. The structure has a cylindrical cross-section and constant density. The reference and the deformed configurations are presented in Figure 4. For sake of simplicity, it is assumed that, in the reference configuration, the line of centroids is a straight line and that the base vector $\boldsymbol{E_3}$, describing the direction of the line of centroids, coincides with the $X_3$ axis of the inertial reference frame. The other two base vectors $\boldsymbol{E_1}$ and $\boldsymbol{E_2}$ are directed along the principal axis of inertia of the cross-section. The position vector $\boldsymbol{D_o}$ of a material particle $P_0$ in the reference configuration can be written as:

$$\boldsymbol{D_o} = \boldsymbol{d_o}(X3) + \boldsymbol{R_o}(X1, X2) \qquad (11)$$

In the deformed configuration the position vector $\boldsymbol{D}$ of the same material particle $P$ can be written as:

$$\boldsymbol{D} = [\boldsymbol{d_o}(X3) + \boldsymbol{u}(X_3)] + \boldsymbol{R}(X1, X2) \qquad (12)$$

where $\boldsymbol{R}$ is the radius vector in deformed configuration. Because of the assumption of rigid cross-sections the vector $\boldsymbol{R}$ in deformed configuration can be seen as a rigid rotation of the vector $\boldsymbol{R_o}$ in the reference configuration:

$$\boldsymbol{D} = [\boldsymbol{d_o}(X3) + \boldsymbol{u}(X_3)] + \boldsymbol{\Lambda}(X_3)\boldsymbol{R_o}(X1, X2) \qquad (13)$$

where $\boldsymbol{\Lambda}$ is a rotation matrix belonging to the $3D$ rotation group $SO(3)$ [23]. Following this theory, at each instant the deformed configuration of the beam is defined by means of the position vector $\boldsymbol{d}$, which defines the deformed configuration of the line of centroids, and by means of a rigid rotation defined by the rotation matrix $\Lambda$. The rotation matrix can be described by means of the rotation vector $\boldsymbol{\varphi} = [\varphi_1 \; \varphi_2 \; \varphi_3]^T$ that has the same direction of the axis of rotation and its modulus is equal to the angle of the rotation. The relationship between the rotation vector and the rotation matrix can be written as:

$$\boldsymbol{\Lambda} = \exp\tilde{\boldsymbol{\varphi}} = \boldsymbol{I} + \frac{\sin\|\boldsymbol{\varphi}\|}{\|\boldsymbol{\varphi}\|}\tilde{\boldsymbol{\varphi}} + \frac{1-\cos\|\boldsymbol{\varphi}\|}{\boldsymbol{\varphi}\cdot\boldsymbol{\varphi}}\tilde{\boldsymbol{\varphi}}^2 \qquad (14)$$

where $\tilde{\boldsymbol{\varphi}}$ is the skew symmetric matrix associated with the rotation vector. Each cross-section of the beam is described by 6 degrees of freedom, 3 translations and 3 rotations. It is important to underline that, contrary to translations, which belong to a vector space, rotations cannot be updated using a simple summation. The updating procedure between two consecutive time steps can be written using the following expression:

$$\boldsymbol{d_{n+1}} = \boldsymbol{d_n} + \boldsymbol{u_n} \qquad (15)$$

$$\boldsymbol{\Lambda_{n+1}} = \exp\tilde{\phi}_n \boldsymbol{\Lambda_n} \qquad (16)$$

Where $\tilde{\phi}$ is the pseudo-rotation vector that transform the basis $\boldsymbol{G_{i,n}}$ into the basis $\boldsymbol{G_{i,n+1}}$. The subscript $n$ denotes the temporal discrete of a time-varying quantity at time $t_n$. Here only the kinematic assumption and the update procedure are recalled because they are useful to explain the coupling between the fluid and the structural problem. For more details about the FEM discretisation one may refer to [46].

*4.3. Setting of the coupled problem*

As mentioned in previous sections. The coupled problem is solved using a partitioned approach. The time step of the coupled simulation is the same as the structural and the fluid computation. Since a strong coupling algorithm is used, an iteration procedure is needed at each time step. A block Gauss-Seidel algorithm is used to solve the iteration procedure and an Aitken's relaxation technique is used to speed up the computation. For each time step at the beginning of the iteration procedure the initial value of the relaxation value is equal to 0.7; for successive iterations the relaxation factor is evaluated using the Aitken's formula [31]. Once the two sub-problems are defined and Algorithm 1 is introduced one needs to develop a mesh motion algorithm and to implement an efficient DFMT method. The development of a DFTM algorithm is particularly delicate because one needs to pay attention to develop a procedure which is energy-conserving. This procedure is not always simple, usually in FSI problems one has to deal with non-matching meshes, and in this particular case even different spatial dimensions are involved. In fact, the beam is a mono-dimensional object, while the fluid forces, obtained via pressure integration and linear interpolation, are evaluated on the FSI interface, which is a 2-D surface in three-dimensional space. For each time step of the coupling algorithm one needs to update the mesh of the fluid problem using the displacements coming from the solid computation, and one needs to transfer fluid forces evaluated on the FSI interface to the nodes of the beam element. The problem is solved making the hypotheses of non-deformable cross-sections. Fluid forces are obtained by circumferential integration and transferred to the beam nodes through the shape functions of the beam element. Fluid forces may be not always orthogonal to the FSI interface because of viscous stresses. For this reason, both forces and moments need to be transferred. At the beginning of each simulation one has to define a mapping between the nodes on the FSI interface and the beam elements, and vice versa. In other words, one has to assign to each beam element a certain number of nodes which insists on it. This procedure is performed by only making geometrical considerations on the initial in-deformed geometry. Since the assumption of in-deformable cross-section is done, this mapping also remains the same during the beam deformation. Once this mapping is defined, focusing only on one beam element, is possible to write the procedure to transform forces acting on the FSI interface into forces and moments acting on the nodes of the beam (Figure 5):

$$\boldsymbol{F_{sj}} = \sum_{i=1}^{Nf} I_j(\xi_i) \boldsymbol{F_{f_i}} \tag{17}$$

$$\boldsymbol{M_{sj}} = \sum_{i=1}^{Nf} I_j(\xi_i) (\boldsymbol{r_{si}} \times \boldsymbol{F_{f_i}}) \tag{18}$$

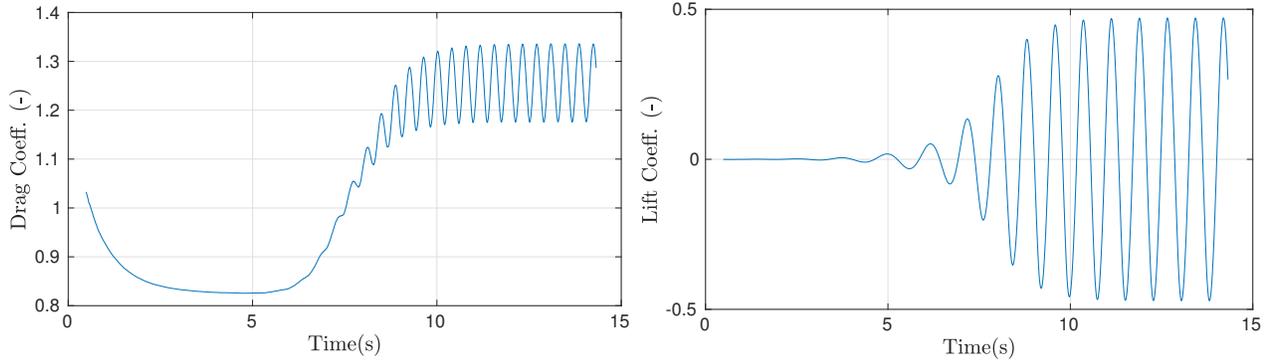

Figure 3: Time evolution of drag and lift coefficients for $Re = 6.31 \times 10^4$. The root mean square of the lift coefficient is equal to 0.33 the mean of the drag coefficient is equal to 1.25. The strohual number is equal to 0.20. The mean of the lift coefficient is equal to 0.0036. For a comparison with reference values see Table 1.

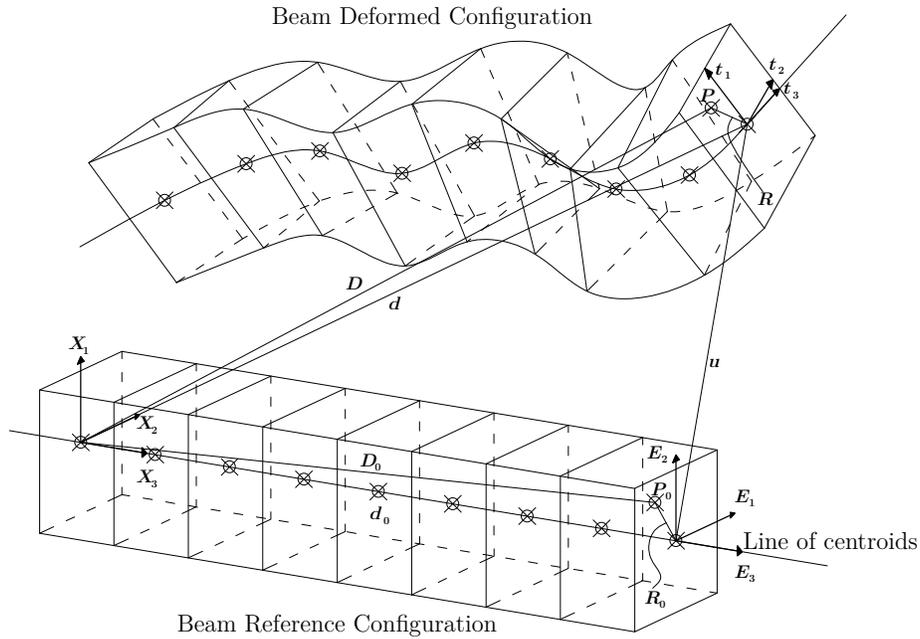

Figure 4: Three dimensional representation of the kinematic assumption used to model the beam element. The image shows the beam reference configuration (in the lower part) with global reference system $\{X_1, X_2, X_3\}$, reference frame $\{E_1, E_2, E_3\}$, reference axis position vector $d_0$, reference cross-section position vector $R_0$, reference position vector $D_0$; The beam deformed configuration (in the upper part) with moving frame $\{t_1, t_2, t_3\}$, deformed axis position vector $d$, deformed position vector $D$ and deformed cross-section position vector $R$.

Where $\boldsymbol{Fs}_j$ is the force acting on the $j^{th}$ node of the beam element, $I_j(\xi_i)$ is the shape function relative to the $j^{th}$ node of the beam evaluated on the natural coordinate $\xi_i$ which correspond to the fluid force $\boldsymbol{F_{f_i}}$. $\boldsymbol{r_{si}}$ is the radius between the $i^{th}$ fluid node and the beam element and $Nf$ is the number of fluid nodes that insists on a certain structural node. The adjoint procedure has to be performed on displacements. Beam nodal displacements have to be properly converted into displacements of the FSI interface. Since an element with finite rotation is involved, and interpolation of rotation variables only makes sense on incremental quantities [26], the updating procedure of the position of the FSI interface has to be performed in an incremental way:

$$\boldsymbol{u_{f_{i,n+1}}} = \boldsymbol{u_{f_{i,n}}} + \sum_{j=1}^{2} \left[ I_j(\xi_i) \Delta \boldsymbol{u}_j + \boldsymbol{\Lambda}(I_j(\xi_i) \Delta \boldsymbol{\varphi}_j) \boldsymbol{r_{si,n}} \right] \tag{19}$$

where $\boldsymbol{u_{f_{i,n+1}}}$ and $\boldsymbol{u_{f_{i,n}}}$ are the displacement of the $i^{th}$ fluid node on the FSI interface respectively at time $t_{n+1}$ and $t_n$, $\Delta \boldsymbol{u}_j$ and $\Delta \boldsymbol{\varphi}_j$ are the incremental displacement and rotation vector, and $\boldsymbol{r_{si,n}}$ is the radius between the $i^{th}$ fluid node on the FSI interface and the relative beam at time $t_n$.

The coupling of the fluid and the solid problems, as already mentioned in § 4, is realized in a strong way using a block Gauss-Seidel coupling algorithm. For the sake of simplicity, and because of the limited computational cost of the solid problem, that only consists of a limited number of beam elements, the time step of the fluid problem is imposed to be equal to the time step of the solid problem. Fluid points that do not belong to the FSI interface are moved according to a Laplacian smoothing algorithm [25]: the equation of cell motion is solved based on the Laplacian of the diffusivity and the cell displacements. The diffusivity field is quadratically based on the inverse of the cell centre distance to the FSI interface.

## 5. The Reduced Order Model

The high fidelity solver discussed in § 4 is used to produce computational data useful for the identification of a reduced order model as highlighted in Figure 1. In particular at the ends of the small scale high fidelity model it is imposed an input motion that matches the expected response in full scale. In particular the reduced order model introduced in this paper consists in the combination of a forced van der Pol wake oscillator model, along the cross flow direction, and a forced linear state-space model, along the in-line direction. The model used is reported in Equations 20 and 21.

$$\dot{\boldsymbol{x}}(t) = \boldsymbol{A}\boldsymbol{x}(t) + \boldsymbol{B}f(L_c(t)) \tag{20}$$
$$D'_c(t) = \boldsymbol{C}\boldsymbol{x}(t) + \boldsymbol{D}f(L_c(t))$$

$$\begin{pmatrix} \dot{x}_{CF,1} \\ \dot{x}_{CF,2} \end{pmatrix} = \begin{pmatrix} x_{2,CF} \\ \mu_{CF}(A_{CF} - x_{CF,1}^2)x_{CF,2} - \omega_{0,CF}^2 x_{CF,1} + B_{CF}f(d) \end{pmatrix}$$
$$L_c = x_{CF,1} \tag{21}$$

In the proposed model, the fluctuating part of the drag coefficient $D'_c = D_c - D_{c,m}$ is modelled using a linear state-space model which uses a function of the lift coefficient as forcing term. In the state space model of Equation 20, $x(t)$ is the state vector of dimension $n \times 1$, $y(t)$ is the output vector of dimension $q \times 1$, $f(L_c(t))$ which is a function of the lift coefficient corresponds to the input vector of dimension $p \times 1$, $A$ is the state matrix of dimension $n \times n$, $B$ is the input matrix of dimension $n \times p$, $C$ is the output matrix of dimension $q \times n$, $D$ is the feed-through matrix of dimension $q \times p$. In this particular case $p = q = 1$ and $n$ depends on the particular case. The size of the state vector and the matrices of the systems need to be determined following the procedure highlighted in subsection 5.1.

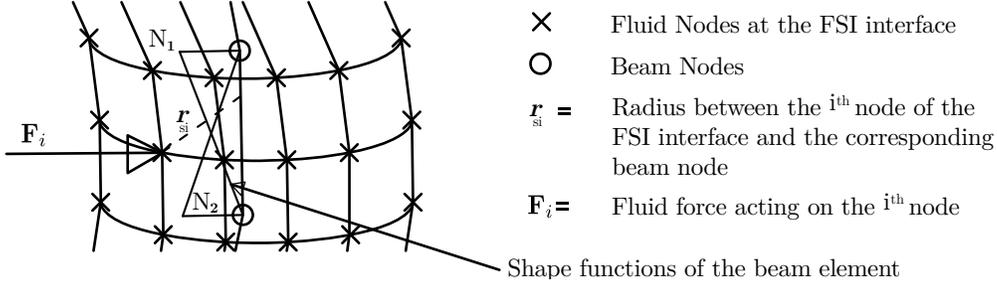

Figure 5: The figure shows the strategy used for the coupling between the fluid and the structural mesh.

Instead, the lift coefficient is modelled using a VDP system which is forced using a function of the instantaneous motion of the cylinder along the CF direction. where $x_{CF}$ is a non-linear state-space vector with two components. In particular $x_{CF,1} = L_c$ and $x_{CF,2} = \dot{L}_c$. $u_{CF}$ is the forcing term of the wake oscillator equation; in theory it may be any physical quantity but it is evident that it is influenced by the motion of the structure within the flow. $\mu_{CF}$, $A_{CF}$, $\omega_{0,CF}^2$ and $B_{CF}$ are unknown parameters which need to be determined using the procedure of subsection 5.2.

It is important to highlight some differences with previous models, in the model proposed by [15] the total hydrodynamic lift coefficient is decomposed into different terms, one related to the added mass effect and one the vortex lift, in their model only the vortex part of the lift coefficient is modelled and the added mass contribution is added in the structural model.

$$(m_s + m_f)\ddot{d}_{CF} + (c_s + c_f)\dot{d}_{CF} + k_s d_{CF} = F_{VORTEX} \quad (22)$$

$$m_s \ddot{d}_{CF} + c_s \dot{d}_{CF} + k_s d_{CF} = F_{TOTAL} \quad (23)$$

Here it is modelled directly the total lift coefficient for this reason one does not need to introduce any added mass term in the structural model. It is chosen to force the state-space model which reproduces the drag coefficient whit a function of the lift coefficient to establish a coupling between the two directions. The turbulence is indeed a threedimensional phenomenon for this reason the two directions must be coupled together.

### 5.1. Identification of the model along the IL direction

To identify the matrices of Equation 20 it is used an iterative approach that uses a prediction error minimization (PEM) algorithm. PEM methods here are briefly summarized from [35]. The idea behind the method is to find a set of parameters $\theta$ such that the prediction error $e(t, \theta)$ is minimized:

$$e(t, \theta) = y(t) - \hat{y}(t|t-1; \theta) \quad (24)$$

where $\theta$ contains the parameters of the model, $t$ is the time, $y(t)$ is the measured output obtained with the high fidelity solver and $\hat{y}(t|t-1; \theta)$ is the prediction of $y(t)$ given the data up to $t - 1$ and the parameters $\theta$. Once the measured input and output are known, and once the model structure is fixed, one can evaluate the parameters by minimization of:

$$\hat{\theta} = \underset{\theta}{\operatorname{argmin}} \frac{1}{n} \sum_{t=1}^{n} F(e(t, \theta)) \quad (25)$$

where $F$ is a properly chosen loss function. In this particular case the loss function is simply $F(e) = e^2$. This method works as long as the dimension of the state vector is known. The size of the state vector is a priori normally not known and one should perform the identification procedure using different sizes of the state vector in order to obtain the optimal size. In order to avoid this trial and error procedure the optimal size of the state vector can be derived by looking at eigenvalues of the Hankel matrix. The only thing one knows a priori are the input and the output data of the system. Using the input and output data the Markov parameters can be identified following the procedure presented in [27] and [39] and become the entries of the Hankel matrix. Once the Markov parameters are determined one can construct the Hankel matrix using the procedure reported in [27]. Applying a singular value decomposition to the Hankel matrix and looking to the eigenvalues one can understand the optimal order of the system. The rank of the Hankel matrix is in fact related to the order of the system. In case of full rank matrix which is often the case with real measurements data, checking the dimension of the eigenvalues one can discard all the eigenvalues smaller than a certain value.

### 5.2. Identification of the model along the CF direction

The parameters of Equation 21 are determined using a non-linear least-squares solver as presented [12, 11]. If all the parameters are collected in a vector $\boldsymbol{p}$ the parameter identification problem can be rewritten as:

$$\min_{\boldsymbol{p}} \frac{1}{2}||\boldsymbol{R}(\boldsymbol{p})||_2^2 = \frac{1}{2}\sum_i r_i(\boldsymbol{p})^2 \qquad (26)$$

where $\boldsymbol{R}(\boldsymbol{p})$ is a function which returns a vector value. In this case $\boldsymbol{R}(\boldsymbol{p})$ is the residual function which returns the difference between the high-fidelity and the ROM output:

$$\boldsymbol{R}(\boldsymbol{p}) = \boldsymbol{F_c}(t) - \boldsymbol{y}(\boldsymbol{p}, t) \qquad (27)$$

where $\boldsymbol{y}$ is the output of the non-linear system and $t$ is the time. This optimization problem is solved herein using a trust region reflective algorithm as presented in [12, 11].

### 5.3. Selection of the forcing terms for the ROM

So far it is not specified the functions used to force the model along the IL and CF directions. It is only mentioned that along the IL direction a function of the lift coefficient is used and along the CF direction a function of the istantaneous motion is used. In this subsection it is discussed the selection of the forcing terms used for the reduced order model. To identify the parameters and the matrices of the system reported in Eq. 20 and 21 we need to force the high fidelety model with an input motion that is resembling the full scale behaviour. The best way would be using full-scale measurements that unfortunately in many cases are not available and since the accuracy of the proposed method is based on the input motion used to force the small-scale model it is necessary to find an accurate way to predict the expected displacements in full-scale. Without any information, a possible way could be to use a white noise that would excite all the possible frequencies of the system. However, this approach is computationally expensive. A simulation time which is long enough to cover all the possible physical frequencies of the system has to be used. Moreover, this approach may produce non-physical and high frequencies that may lead the FSI solver to diverge. The statistical properties of the imposed motion are also obtained through high-fidelity simulations on a small-scale size. The cable is retained at both ends with suitable springs that replace the remaining parts of the cable. The displacements along the IL and CF directions are measured in the middle point of the structure and used to force another high fidelity simulation that is later used for the identification. In particular the high-fidelity model of § 4 is restrained at both end with springs following the setting of Figure 6. The values of the stiffness properties of the springs have been deduced investigating the FEM structural model of a steel full scale cable with cylindrical cross section.

#### 5.3.1. Forcing term along the IL direction

[38] modelled the unsteady force coefficients on a stationary cylinder using a VDP model for the lift coefficient and a term proportional to $L_c \cdot \dot{L}_c$ for the drag coefficient.

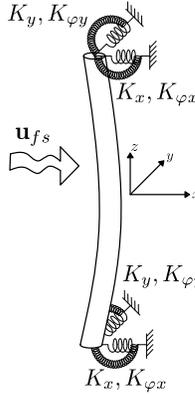

| Par. | Value | unit |
|---|---|---|
| $E_s$ | $5.88 \times 10^8$ | N/m$^2$ |
| $D$ | 0.102 | m |
| $\rho_s$ | $5.582 \times 10^3$ | kg/m$^3$ |
| $\nu_s$ | 0.3 | (-) |
| $L_s$ | 40D | m |
| $K_x$ | 576 | m |
| $K_{\varphi x}$ | 803 | Nm/rad |
| $K_y$ | 409 | N/m |
| $K_{\varphi y}$ | 10417 | Nm/rad |
| $\boldsymbol{u}_{fs}$ | 0.5 | m/s |

Figure 6: The figure and the table are showing the setting of the full order problem used to gather numerical results for the identification performed in the following sections. $E_s$ is the Young's modulus, $D$ is the diameter, $\rho_s$ is the density of the cable's material, $\nu_s$ is the Poisson's ratio, $L_s$ is the length of the cable, $K_x,K_{\varphi x},K_y,K_{\varphi y}$ are the stiffness properties of the springs and $\boldsymbol{u}_{fs}$ is the free stream velocity.

[40] stated that the linear relationship should be proportional to the term $L_c^2$ while [2] concluded that both terms contribute to the modelling of the lift. Since there is no universal agreement in literature, especially dealing with moving cylinders, here both models are proposed substituting the function $f(L_c(t))$ with $L_c^2(t)$ and $L_c(t) \cdot \dot{L}_c(t)$ respectively. The results of the identified models are proposed in Figure 7. In this case, the model which gives the best results is the one which uses $L_C^2$ as input function of the state-space model. The appropriate size of the model turns out to be equal to 8. The comparison between the two models is reported in the time domain in Figure 7 and in terms of best fit in Table 2. The best fit is evaluated as:

$$\text{B.F.} = \left(1 - \frac{|z - \hat{z}|}{|z - \overline{z}|}\right) \cdot 100 \qquad (28)$$

where $z$ is the measured output, $\hat{z}$ is the simulated or predicted model output and $\overline{z}$ is the mean of the measured output. The two considered model used to approximate the in-line forces are:

$$\dot{\boldsymbol{x}}(t) = \boldsymbol{A}\boldsymbol{x}(t) + \boldsymbol{B}L_c^2(t) \qquad (29)$$
$$D'_c(t) = \boldsymbol{C}\boldsymbol{x}(t) + \boldsymbol{D}L_c^2(t)$$
$$\dot{\boldsymbol{x}}(t) = \boldsymbol{A}\boldsymbol{x}(t) + \boldsymbol{B}L_c(t)\cdot\dot{L}_c(t) \qquad (30)$$
$$D'_c(t) = \boldsymbol{C}\boldsymbol{x}(t) + \boldsymbol{D}L_c(t)\cdot\dot{L}_c(t)$$

Where the matrices $\boldsymbol{A}$, $\boldsymbol{B}$, $\boldsymbol{C}$, $\boldsymbol{D}$ need to be identified using the procedure reported in subsection 5.1, $D'_c$ is the fluctuating part of the drag coefficient, $L_c$ and $\dot{L}_c$ are the lift coefficient and its time derivative respectively.

#### 5.3.2. Forcing term along the CF direction

Regarding the forcing term for the wake oscillator equation which models the lift coefficient, three different model have been tested which dependes linearly respectively to the displacement, velocity and acceleration of the cylinder along the CF direction. It means that inside Equation 21 $u_{CF} = d_{CF}, \dot{d}_{CF}, \ddot{d}_{CF}$, where $d_{CF}$ is the cross flow displacement of the cylinder at the ends. The three different models considered to model the cross-flow forces are then:

$$\begin{pmatrix} \dot{x}_{CF,1} \\ \dot{x}_{CF,2} \end{pmatrix} = \begin{pmatrix} x_{2,CF} \\ \mu_{CF}(A_{CF} - x_{CF,1}^2)x_{CF,2} - \omega_{0,CF}^2 x_{CF,1} + B_{CF} d_{CF} \end{pmatrix}$$
$$L_c = x_{CF,1} \quad (31)$$

$$\begin{pmatrix} \dot{x}_{CF,1} \\ \dot{x}_{CF,2} \end{pmatrix} = \begin{pmatrix} x_{2,CF} \\ \mu_{CF}(A_{CF} - x_{CF,1}^2)x_{CF,2} - \omega_{0,CF}^2 x_{CF,1} + B_{CF} \dot{d}_{CF} \end{pmatrix}$$
$$L_c = x_{CF,1} \quad (32)$$

$$\begin{pmatrix} \dot{x}_{CF,1} \\ \dot{x}_{CF,2} \end{pmatrix} = \begin{pmatrix} x_{2,CF} \\ \mu_{CF}(A_{CF} - x_{CF,1}^2)x_{CF,2} - \omega_{0,CF}^2 x_{CF,1} + B_{CF} \ddot{d}_{CF} \end{pmatrix}$$
$$L_c = x_{CF,1} \quad (33)$$

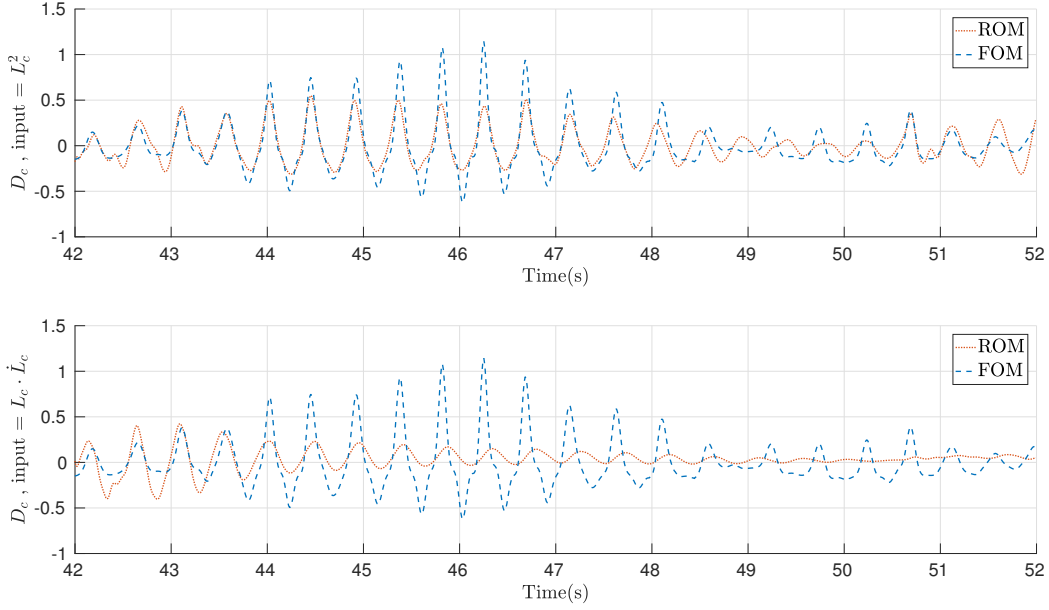

Figure 7: Comparison of the drag coefficient obtained with the two different reduced order models of Equations 29 and 30. In both graphs the response using the ROM is indicated with a dotted red line while the full order model (FOM) response is indicated with a dashed blue line. In the top graph are shown the results for the ROM indicated in Equation 29. The lower graph shows the results of the ROM indicated in Equation 30. The FOM response is obtained using the setting of Figure 6.

| Name | Best Fit |
|---|---|
| $L_c^2$ | 46.49 % |
| $L_c \cdot \dot{L}_c$ | 8.74 % |

Table 2: Best fit values for the two proposed ROMs used to model the in-line forces. The plot of the two different models are shown in Figure 7. The best fit values are computed using the discrete $L^2$ norm.

Where the parameters $\mu_{CF}$, $A_{CF}$, $\omega_{0,CF}^2$ and $B_{CF}$ need to be identified following the procedure of subsection 5.2, $L_c$ is the lift coefficient and $d_{CF}$, $\dot{d}_{CF}$, $\ddot{d}_{CF}$ are the cross flow displacement and its first and second order time derivatives. In literature, these models are used mainly to describe the cross-flow motion of 1 degree of freedom rigid cylinders. [15] studied these three models, concluding that the coupling model that gives the best results is the acceleration coupling model. The results, in terms of best fit, obtained for the three different models are reported in Table 3. As can be seen from the table the acceleration-coupling model gives the best fit. This result is in accordance with what discovered by [15]. The comparison between the measured and the simulated outputs is reported in time domain in Figure 8 and in frequency domain in Figure 9 for displacement, velocity and acceleration-coupling respectively.

| Name | Best Fit |
|---|---|
| CF LC dis. | 33.23 % |
| CF LC vel. | 12.92 % |
| CF LC acc. | 34.12 % |

Table 3: Best fit values for three different ROMs used to model the cross flow (CF) forces. The best fit refers to Figure 8 and is calculated using the lift coefficient (LC) obtained using the FOM model and the one computed with three different ROMs of Equations 31, 32 and 33. The FOM response is evaluated using the setting of Figure 6. The best fit values are computed using the discrete $L^2$ norm.

## 6. Numerical results using experimental investigations

The comparison performed in subsubsection 5.3.1 and subsubsection 5.3.2 gives some informations about the

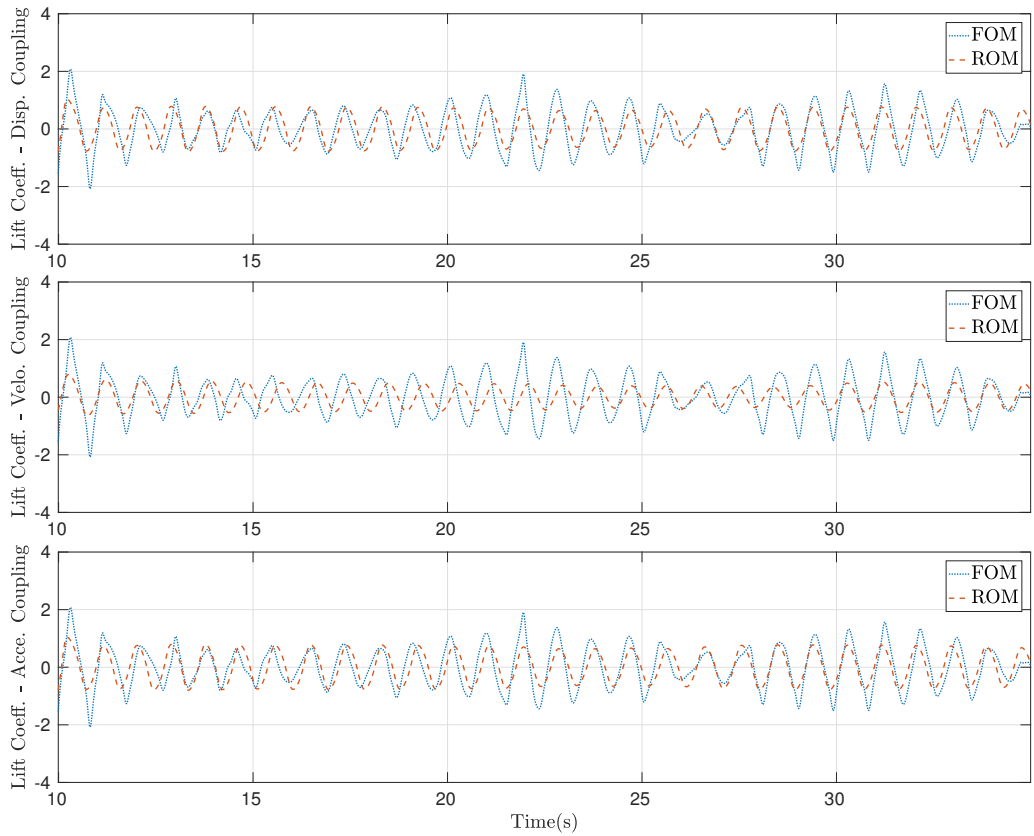

Figure 8: Comparison in the time domain of the lift coefficient obtained with three different reduced order models. In all the graphs the response using the ROM is indicated with a dashed red line while the full order model (FOM) response is indicated with a dotted blue line. From the top the bottom are shown the graphs with the displacement, velocity and acceleration coupling model presented in Equations 31, 32 and 33 respectively. The FOM response is obtained using the setting of Figure 6.

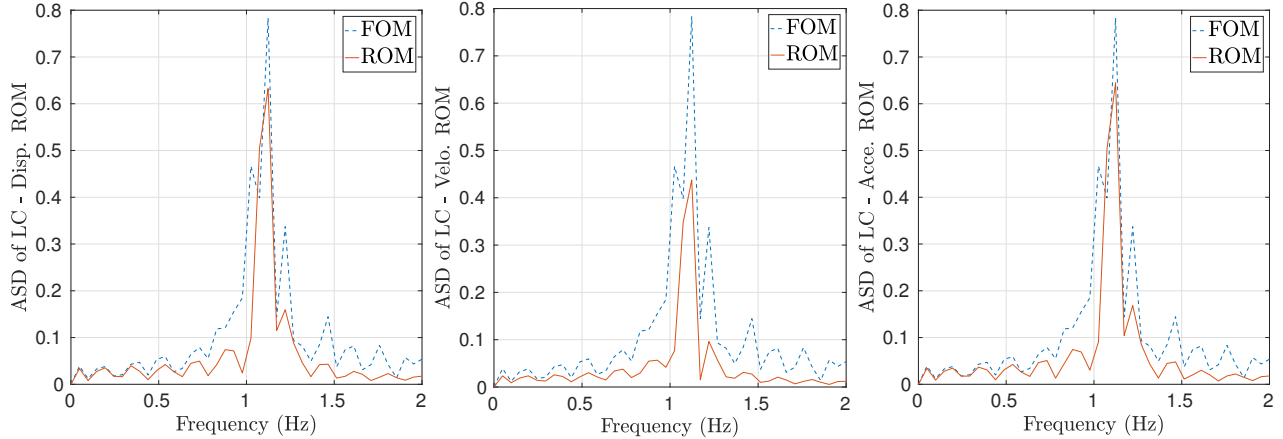

Figure 9: Comparison in the frequency domain of the amplitude spectral density (ASD) of the lift coefficient obtained with three different reduced order models. In all the graphs the response using the ROM is indicated with a continuous red line while the full order model (FOM) response is indicated with a dashed blue line. From right to left are shown the graphs with the displacement, velocity and acceleration coupling model presented in Equations 31, 32 and 33 respectively. The FOM response is obtained using the setting of Figure 6.

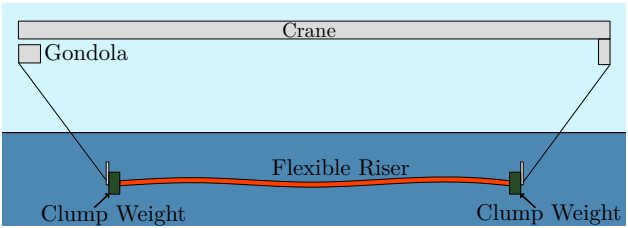

Figure 10: Setting of the NDP 38 m experiment setting. Image adapted from [53]

quality of the identified model but does not give any information regarding the behaviour of the identified model in full-scale when one tries to use the identified model to force a full-scale structure. In order to validate the developed model, measurements from full-scale experiments are used, particularly the data set from experimental investigations conducted by the Norwegian Marine Technology Research Institute (Marintek) inside their towing tank located in Trondheim. The set up of the experimental campaign is depicted in Figure 10. A 38m long riser is tested. The riser is free to move along the in-line and the cross-flow directions; only the torsional degree of freedom is constrained. Clump weights acting in tandem with pendulum arms, and placed at the ends, are used to moderate the tension variation along the riser. Uniform and linearly sheared flow velocity profiles are investigated. The uniform flow velocity profile is generated by towing the riser with a constant velocity along the towing tank while the linearly sheared flow is generated by a rotation of the crane around its vertical axis. Different riser types are tested: a bare riser and a riser partially or totally covered with helical strakes to investigate their effect on VIV suppression. Here the focus is only on the bare type. The mechanical and geometrical properties of the riser are reported in Table 4. The tension is varying along the riser during the tests but for the validation a constant tension is applied to the full-scale finite element model. Only the main features of the test setting are reported here - for more details see [53]. The instrumentation placed along the riser counts

| Quantity | Value | Unit |
|---|---|---|
| $D$ | 0.027 | m |
| $E$ | $8.894 \cdot 10^8$ | $\frac{N}{m^2}$ |
| $A$ | $5.7 \cdot 10^{-4}$ | $m^2$ |
| $J$ | $4.2 \cdot 10^{-8}$ | $m^4$ |
| $\rho$ | $1.630 \cdot 10^3$ | $\frac{kg}{m^3}$ |
| length ($l$) | 38 | m |
| tension ($T$) | 3000 | N |

Table 4: Structural characteristics of the riser model adopted in the experimental investigations and used for the validation of the proposed model.

24 strain gauges and 8 accelerometers along the CF direction and 40 strain gauges and 8 accelerometers along the IL direction. The sampling rate is equal to 1200Hz. The accelerometers are placed at different positions along the riser length, particularly at $z = 5m$, 10m, 14m, 19m, 22m, 26m, 31m and 36m, where $z$ is the axis along the length of the riser. Several flow velocities are tested in experimental conditions but only velocities equal to 1.4m/s and 1.7m/s are available to the author of this paper. The time histories of the accelerations measured in the experiments are used to force the high-fidelity small-scale model and the force coefficients are measured.

The output of the developed high-fidelity simulations are used together with the time histories of motion to feed the system/parameter identification technique presented in § 5, and the parameters and the system size of model presented in Equation 20 and 21 are determined. The developed model is then used to force the global finite element model of the structure and the results are compared in terms of displacements with the results of the experi-ments.

*6.0.1. Flow velocity* $\mathbf{u}_{fs} = 1.4\ m/s$

The first comparison is performed in small-scale in order to check whether the model is also able to perform a comparison in full-scale. The comparison is performed both in time and frequency domains and is reported in Figures 11 and 12. As mentioned in § 5, the identi-fied model is constructed using a linear state-space model forced by a term which depends linearly on the square of the lift coefficient for the IL direction and a forced van der Pol model along the CF direction. The optimal model size turns out to be equal to 6. The non-linear deterministic identification procedure led to the parameters of the VDP model reported in Table 6. The simulation performed us-ing the identified model, especially along the CF direction, matches well with the high-fidelity data.

| Name | Best Fit |
|---|---|
| Drag Coeff. | 20.88 % |
| Lift Coeff. | 65.6 % |

Table 5: Best fit values for the Drag and Lift coefficients. The values refer to Figure 11, the free stream flow velocity is equal to $\boldsymbol{u}_{fs} = 1.4$m/s. The best fit values are computed using the discrete $L^2$ norm.

| Name | Value |
|---|---|
| $\mu_{CF}$ | 68.29 |
| $A_{CF}$ | 1.18 |
| $\omega_{0,CF}^2$ | 2117 |
| $B_{CF}$ | 70.68 |

Table 6: Identified parameters of the VDP model for $\boldsymbol{u}_{fs} = 1.4$m/s

In order to perform the comparison in full-scale the finite element model of the structure is built. The finite element model is constructed using 50 geometrically exact beam elements described in § 4 and matching the mechanical properties reported in Table 4. At the ends of the full-scale model only translational and torsional degrees of freedom are retained. At each node of the structure hydrodynamic forces are modelled using the identified model and displacements are measured. The steady part of the drag is applied as a static load before the beginning of the VIV phenomena. The steady drag matches the mean value measured during the high-fidelity simulation. In this case, the mean value of the drag coefficient is equal to $D_{c,m} = 2.34$. This value is higher than the values normally observed on static cylinders but this is a common feature of oscillating cylinders. The amplification of the mean (steady) drag coefficient was observed by many authors [30, 56]. The comparison is performed in the frequency domain in terms of displacement. It is performed only in the frequency domain because the initial conditions and the small transient may lead to a phase shift between the measured and simulated response. In Figure 13 the comparison of the power spectral density between the experiment and the simulation in terms of displacements is reported. Results are presented for $z = 19$, located at the midpoint of the full-scale model. As can be seen from the graphs, a good agreement between the simulated and measured response is observed. It is interesting to note that not only the main frequency and amplitude of oscillation are captured, but the trajectory of the motion also matches. The phase angle between the measured and simulated response differs slightly.

*6.0.2. Flow velocity* $\mathbf{u}_{fs} = 1.7\ m/s$

The procedure performed in subsubsection 6.0.1 is repeated here for the other available value of flow velocity. This second test is performed in order to check the sensitivity of the proposed model to the flow velocity and to check whether the realisation of a future model which would also consider the flow velocity as a parameter is possible. The same graphs reported in subsubsection 6.0.1 are also reported here for both the small-scale and the full-scale. As can be seen in Table 7, results are similar to the case with a flow velocity of $\boldsymbol{u}_{fs} = 1.4$m/s. The identified parameters of the VDP system are reported in Table 8.

| Name | Best Fit |
|---|---|
| Drag Coeff. | 20.9 % |
| Lift Coeff. | 71.2 % |

Table 7: Best fit values for the Drag and Lift coefficients. The values refer to Figure 15, the free stream flow velocity is equal to $\boldsymbol{u}_{fs} = 1.7$m/s. The best fit values are computed using the discrete $L^2$ norm.

| Name | Value |
|---|---|
| $\mu_{CF}$ | 88.60 |
| $A_{CF}$ | 1.35 |
| $\omega_{0,CF}^2$ | 3540 |
| $B_{CF}$ | 34.94 |

Table 8: Identified parameters of the VDP model for $\boldsymbol{y}_{fs} = 1.7$ m/s

## 7. Conclusions and Outlooks

This paper aimed to present a novel reduced order model for the analysis of long flexible cylinders in an offshore environment with a focus on vortex induced vibrations.
A reduced order model for the analysis of vortex induced vibrations phenomena in both the in-line and cross-flow directions has been developed. Several different models were tested and discussed. In particular, models based on a forced van der Pol oscillator and on a linear state-space

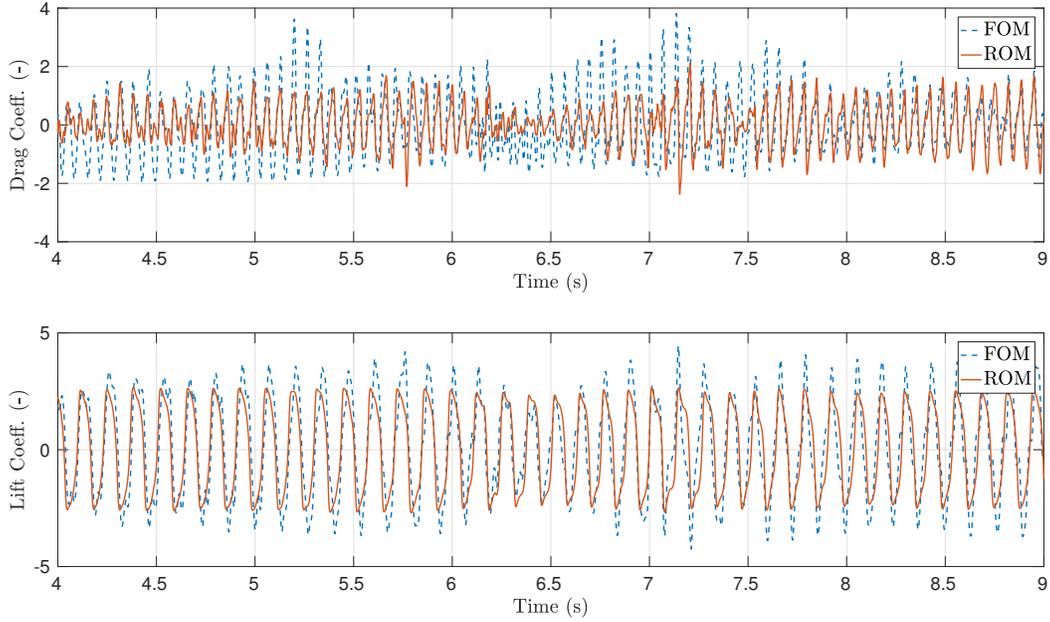

Figure 11: Comparison of the FOM (dashed blue line) and ROM (continuous red line) response in the time domain. In the top graph it is plotted the comparison of the drag coefficient while in the lower graph it is plotted the comparison of the lift coefficient. The FOM force coefficients are obtained using the FSI solver response with the structural properties of Table 4, a lenght $L = 40D$, and an imposed motion at the ends of the cable matching the NDP experiments with inlet velocity equal to $\boldsymbol{u}_{fs} = 1.4$m/s.

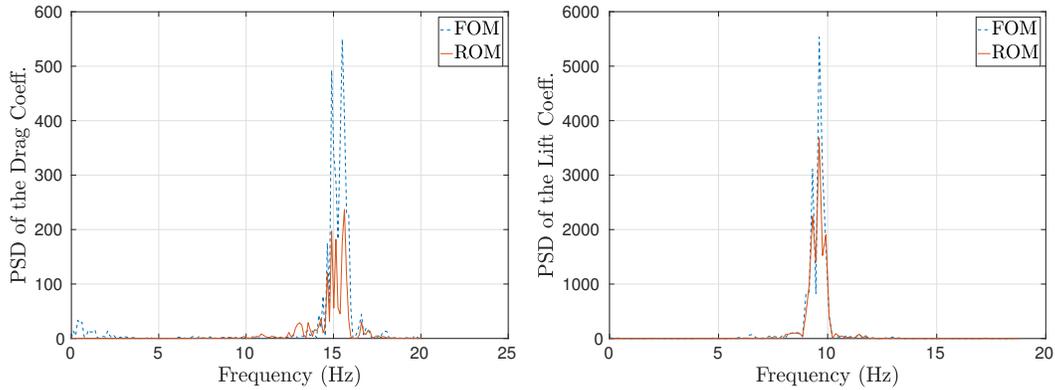

Figure 12: Comparison of the FOM (dashed blue line) and ROM (continuous red line) response in the frequency domain (in terms of power spectral density). In the left graph it is plotted the comparison of the drag coefficient while in the right graph it is plotted the comparison of the lift coefficient. The FOM force coefficients are obtained using the FSI solver response with the structural properties of Table 4, a lenght $L = 40D$, and an imposed motion at the ends of the cable matching the NDP experiments with inlet velocity equal to $\boldsymbol{u}_{fs} = 1.4$m/s.

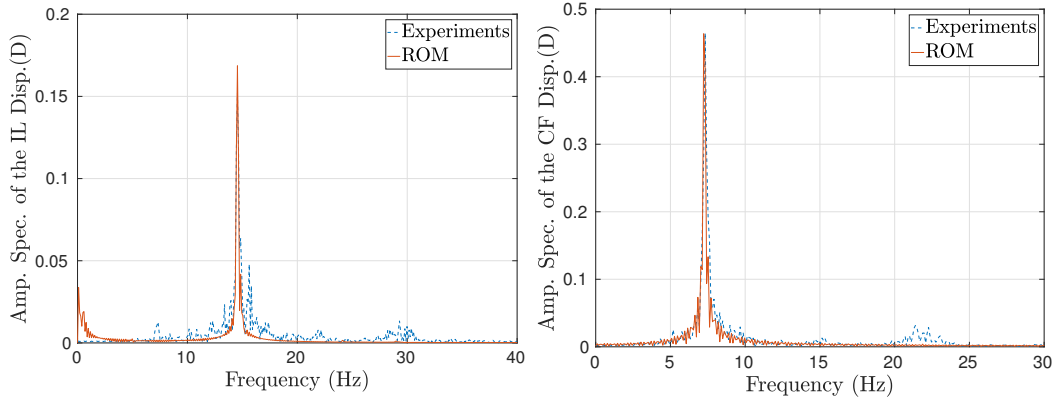

Figure 13: Comparison of results in full scale in terms of displacements in the frequency domain (in terms of amplitude spectral density). The experimental investigations of the NDP dataset (dashed blue line) for $z = 19$m are compared with the numerical values obtained forcing the full scale FEM model of the riser with the identified ROM (continuous red line). The flow velocity is equal to $\boldsymbol{u}_{fs} = 1.4$m/s. On the right is shown the in-line (IL) displacement while on the right is shown the cross-flow (CF) displacement.

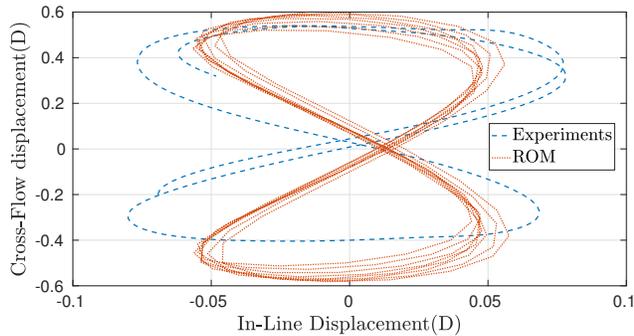

Figure 14: Comparison of the trajectory for $z = 19$ m, $\boldsymbol{u}_{fs} = 1.4$ m/s between the experimental NDP dataset (dashed blue line) and the FEM response forced with the identified ROM (continuos red line)

model are analysed. The final proposed model consists of a forced van der Pol oscillator along the cross-flow direction, and of a linear state space model along the in-line direction. The van der Pol system is forced by using a function of the instantaneous acceleration of the cylinder along the cross-flow direction while the linear state-space system is forced using a function of the square of the lift coefficient. To the author's knowledge, it is the first time that such a model has been used to model vortex induced vibrations phenomena. This reduced order model has the capacity of exploiting the advantages of both the van der Pol equation and of linear state-space models. The van der Pol system has in fact the self-limitation and self-excitation features which are peculiar characteristics of the vortex induced vibrations phenomena. On the other hand, state-space models can model systems with arbitrary dimensions and with more than one dominant frequency. Operating in such a manner the resulting model is self-limited and self-excited for both the in-line and the cross-flow direction. Moreover, in this way, the two directions are coupled.

A fluid-structure interaction solver has been developed by coupling a finite element method solver with a computational fluid dynamic solver with a large eddy simulation turbulence model. The high fidelity solver took approximately two weeks to perform 25s of simulation running in parallel on 20 cores Intel Xeon E5-2680 v2. The results of the fluid-structure interaction analyses, in terms of drag and lift coefficient, conducted on a small scale flexible cylinder, undergoing forced oscillations, are used to identify the parameters and the structure of the reduced order model. The reduced order model, since the number of degrees of freedoms of the fluid part has been extremely reduced, took only approximately one second to perform 25s of simulation running in serial on a laptop with an Intel i7-4710HQ CPU. The coupled system used for the comparison with the experimental investigations, consisting in the full FEM structural model and the fluid ROM used to model the hydrodynamic forces, took approximately 2 minutes to perform 25s running in serial on the same laptop with an Intel i7-4710HQ CPU.

The identified reduced order model has been validated by using experimental investigations conducted in a wave tank showing promising results. The model has the capacity of capturing the amplitude and frequency of oscillation for both the considered flow velocities. The trajectory of the cylinder has also been captured showing that the coupling between the in-line and cross-flow directions is well represented.

Several problems and questions came out. Some of these would need further investigation.

The proposed model was demonstrated to work well for the two analysed flow velocities, but it would be interesting to test the model for a wider range of velocities and to study the possibility of an interpolation algorithm between the different flow velocities.

A factor that is not treated here and would be interesting to study is the inclination between the flow and the cylinder. Only flows that are perpendicular to the axis of the cylinder have been studied and it would be interesting to study the influence of the flow inclination.

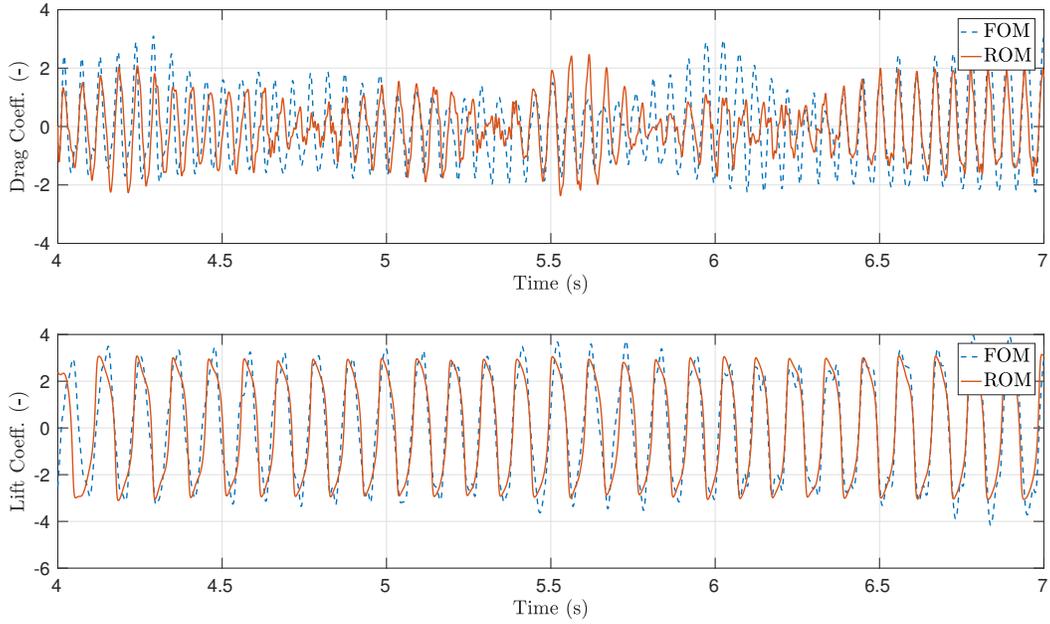

Figure 15: Comparison of the FOM (dashed blue line) and ROM (continuous red line) response in the time domain. In the top graph it is plotted the comparison of the drag coefficient while in the lower graph it is plotted the comparison of the lift coefficient. The FOM force coefficients are obtained using the FSI solver response with the structural properties of Table 4, a lenght $L = 40D$, and an imposed motion at the ends of the cable matching the NDP experiments with inlet velocity equal to $\boldsymbol{u}_{fs} = 1.7$m/s.

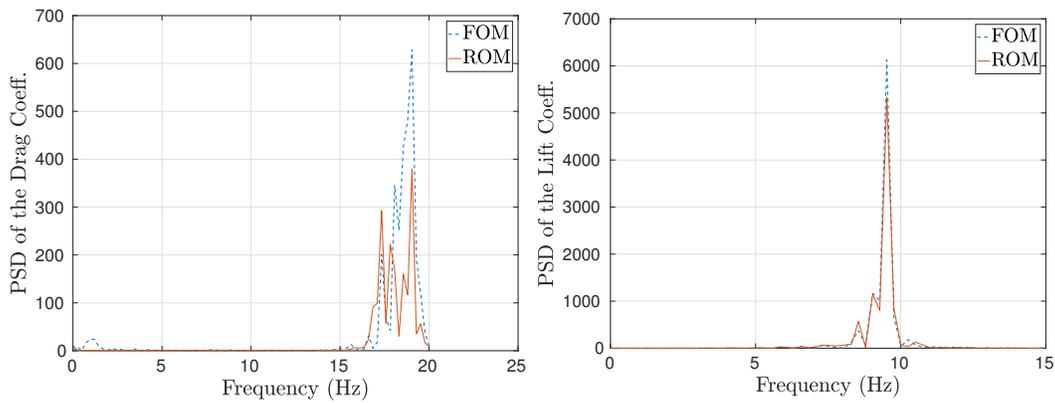

Figure 16: Comparison of the FOM (dashed blue line) and ROM (continuous red line) response in the frequency domain (in terms of power spectral density). In the left graph it is plotted the comparison of the drag coefficient while in the right graph it is plotted the comparison of the lift coefficient. The FOM force coefficients are obtained using the FSI solver response with the structural properties of Table 4, a lenght $L = 40D$, and an imposed motion at the ends of the cable matching the NDP experiments with inlet velocity equal to $\boldsymbol{u}_{fs} = 1.7$m/s.

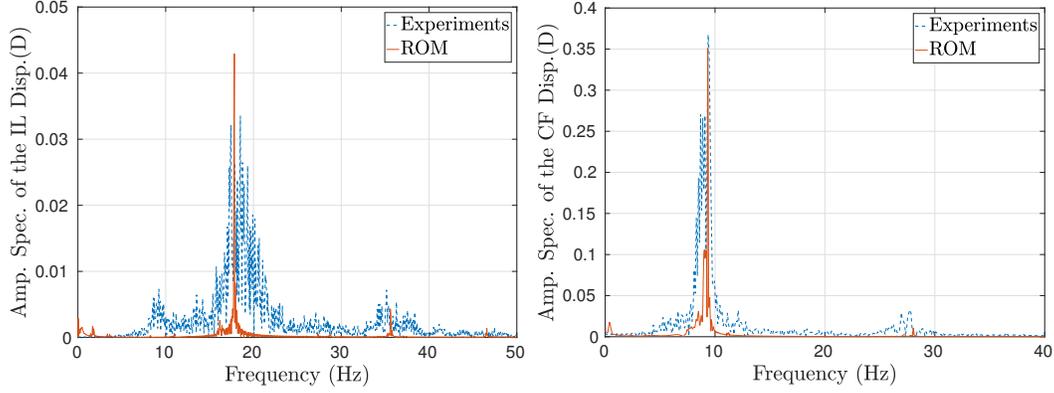

Figure 17: Comparison of results in full scale in terms of displacements in the frequency domain (in terms of amplitude spectral density). The experimental activities of the NDP dataset (dashed blue line) for $z = 19$m are compared with the numerical values obtained forcing the full scale FEM model of the riser with the identified ROM (continuous red line). The flow velocity is equal to $\boldsymbol{u}_{fs} = 1.7$m/s. On the right is left is shown the in-line (IL) displacement while on the right is shown the cross-flow (CF) displacement.

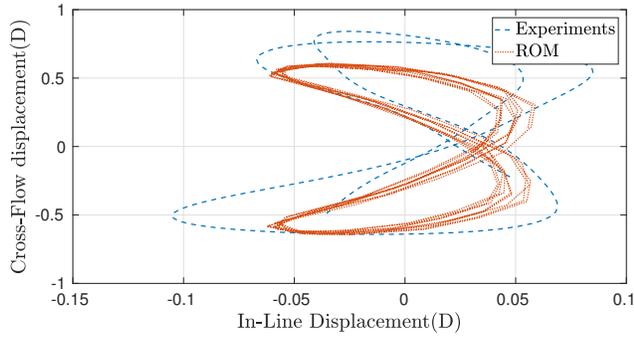

Figure 18: Comparison of the trajectory for $z = 19$ m, $\boldsymbol{u}_{fs} = 1.7$ m/s between the experimental NDP dataset (dashed blue line) and the FEM response forced with the identified ROM (continuos red line).

## 8. Aknowledments

A grateful acknowledgement goes to the Norwegian Deepwater Programme for making available the dataset we used for the validation of this model. We are also grateful to the "Regione Toscana" that financed this work through the "Giovanisí" project and to the DAAD (Deutscher Akademischer Austasch Dienst) institution for the finantial help.

## Appendix A. List of abbreviations and symbols

**Abreviations**

| | |
|---|---|
| ALE | Arbitrary Lagrangian Eulerian |
| CFD | Computational fluid dynamics |
| CTL | Component Template Library |
| DFMT | Direct Force-Motion Transfer |
| DNS | Direct Numerical Simulation |
| FEM | Finite Element Method |
| FSI | Fluid Structure Interaction |
| FVM | Finite Volume Method |
| LES | Large Eddy Simulation |
| PEM | Prediction Error Method |
| POD | Proper Orthogonal Decomposition |
| RANS | Reynolds-averaged Navier-Stokes |
| ROM | Reduced Order Model |
| SISO | Single Input Single Output |
| VDP | van der Pol |
| VIV | Vortex Induced Vibrations |
| $CF$ | Cross-Flow |
| $IL$ | In-Line |

**Symbols**

| | |
|---|---|
| $\boldsymbol{u}_{fs}$ | free stream velocity |
| $D$ | cylinder diameter |
| $d$ | general letter to indicate the displacement of the cylinder |
| $D'_c$ | Fluctuating part of the Drag Coefficient |
| $D_c$ | Total Drag Coefficient |
| $D_{c,m}$ | Mean part of the Drag Coefficient |
| $d_{IL}, d_{CF}$ | displacements along the $IL$ and $CF$ directions |
| $f_v$ | frequency of vortex shedding |
| $L$ | Lenght of the cylinder |
| $L_c$ | Lift Coefficient |
| $Re$ | Reynolds Number |
| $St$ | Strohual Number |